\documentclass[acmlarge, screen, authorversion, nonacm]{acmart}
\AtBeginDocument{%
  \providecommand\BibTeX{{%
    \normalfont B\kern-0.5em{\scshape i\kern-0.25em b}\kern-0.8em\TeX}}}

\usepackage{soul}  

\makeatletter
\g@addto@macro{\UrlBreaks}{\UrlOrds}
\makeatother

\usepackage{subcaption}
\usepackage{url}
\usepackage{listings}       
\usepackage{xcolor}

\begin{document}

\title[]{The Use of Web Archives in Disinformation Research}


\author{Michele C. Weigle}
\orcid{0000-0002-2787-71661234-5678-9012}
\affiliation{%
  \institution{Web Science and Digital Libraries Research Group, Department of Computer Science, Old Dominion University}
  \city{Norfolk}
  \state{VA}
  \country{USA}
  \postcode{23529}
}
\email{mweigle@cs.odu.edu}


\begin{abstract}
In recent years, journalists and other researchers have used web archives as an important resource for their study of disinformation. This paper provides several examples of this use and also brings together some of the work that the Old Dominion University Web Science and Digital Libraries (WS-DL) research group has done in this area.  We will show how web archives have been used to investigate changes to webpages, study archived social media including deleted content, and study known disinformation that has been archived.
\end{abstract}




\maketitle

\section{Introduction}

Web archives have been increasingly used in recent years to study disinformation. A survey \cite{Frew2022Aug} of 100 news articles that mentioned the Internet Archive's Wayback Machine\footnote{\url{https://web.archive.org/}} found that many journalists use web archives to search for deleted content and to determine the veracity of fabricated screenshots. 
Webpages and social media posts can be modified or deleted or authors can be banned from the platform and their posts removed, so one of the first considerations is that we need web archives \cite{weigle-ssrc18} so that we can study the original materials later, even if they contain misinformation.  Mark Graham, Director of the Internet Archive's Wayback Machine, has said, ``It’s not about trying to archive the stuff that’s true, but archive the conversation. All of that is what people are experiencing.'' \cite{ft-IA19} Members of the Election Integrity Partnership noted that during their investigation of the impact of disinformation on the 2020 US Election \cite{eip}, they would archive important pieces of disinformation at the Internet Archive, so that they would be able to go back later and write their report \cite{eip-discussion}. This is becoming a common practice for journalists and investigative reporters, with citations and mentions of web archives, and especially the Wayback Machine, becoming more common in the press. The Internet Archive has even begun keeping track of references to the Wayback Machine in the popular press.\footnote{\url{https://archive.org/about/news-stories/search?mentions-search=Wayback+Machine}} There have been multiple tools and workflows developed to help journalists archive content, such as Bellingcat's auto-archiver \cite{bellingcat-autoarchiver} that can take a list of URLs in a Google Sheet to archive and Raffaele Messuti's extension\footnote{\url{https://docuverse.notion.site/archiveweb-page-brave-tor-939aacd4100c44edb0f7ec5fa4ecc446}} for the Brave browser that uses ArchiveWeb.page\footnote{\url{http://archiveweb.page/}} and Tor\footnote{\url{https://www.torproject.org/}} to allow for anonymous archiving of webpages.

The remainder of this paper, which is an edited version of a post \cite{weigle-blogpost} from the Old Dominion University (ODU) Web Science and Digital Libraries (WS-DL) Research Group blog,\footnote{\url{https://ws-dl.blogspot.com/}} will cover several ways that web archives have been used to study disinformation, including investigating changes to webpages, studying archived social media, studying archived disinformation, and studying archived advertisements.

\section{Investigating Changes to Webpages}

The most straightforward way to use web archives is to investigate changes to a webpage. In late 2022, US Senate candidate Blake Masters was in the news because his campaign removed references to his previous statements on abortion \cite{Roberts2022Aug} and the 2020 Election \cite{Kaczynski2022Aug} from his website. Reporters used the Wayback Machine to compare the text on his live webpage with that from previous versions. Masters was not the only candidate to make such changes heading into the 2022 US midterm general elections, with journalists documenting \cite{AlexiMcCammond2022Aug} several GOP candidates who had ``updated'' their campaign webpages to hide controversial statements and positions. 

\begin{figure}
    \centering
    \includegraphics[width=0.7\textwidth, trim=0 25 0 0, clip]{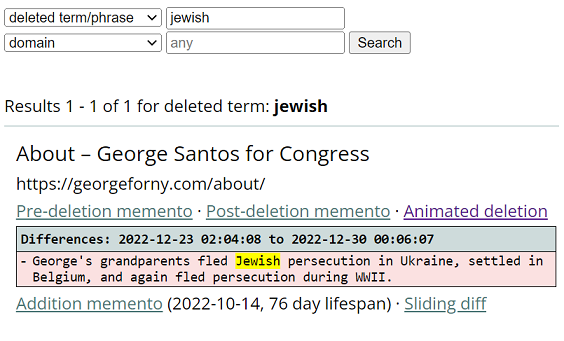}
    \caption{Prototype interface to allow users to search for deleted terms and phrases. The results provide information about the identified changes, links to the mementos, and links to view the change as an animation or a sliding diff interface. \cite{Frew2023Feb}}
    \label{fig:deletion-interface}
\end{figure}

\begin{figure}
    \centering
    \includegraphics[width=0.6\textwidth, trim=0 200 0 0, clip]{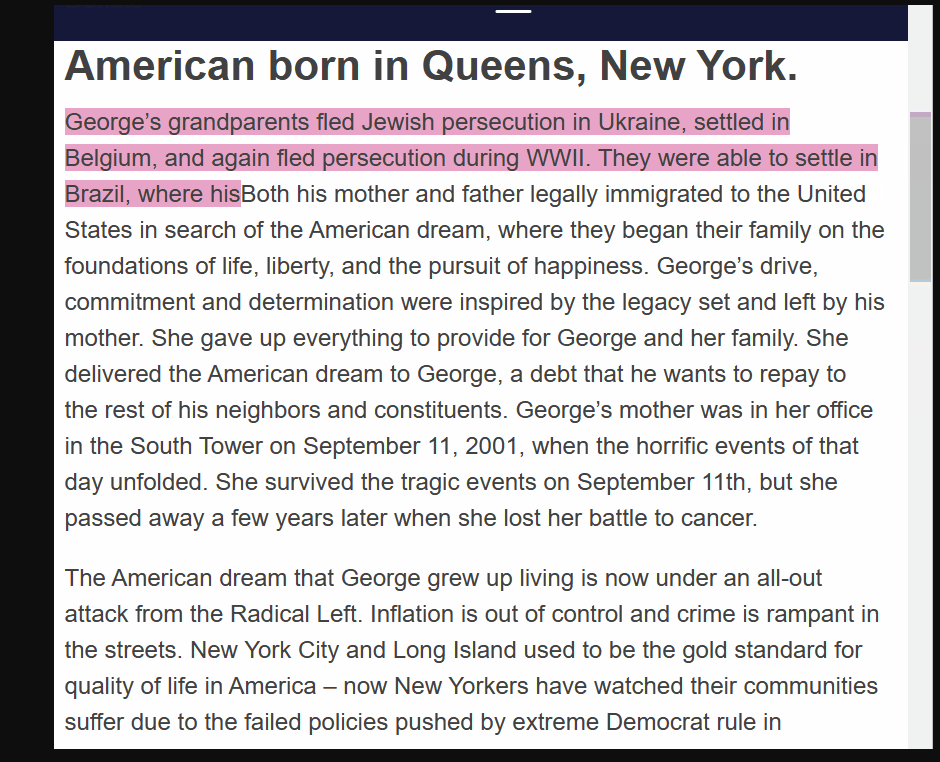}
    \caption{Screenshot from an animation based on the deletion of the statement about George Santos's grandparents being Jewish from his campaign website at \url{https://georgeforny.com/about/}. The text highlighted in red was removed from later versions of the webpage. The animation was created using mementos from \href{https://web.archive.org/web/20221223020408/https://georgeforny.com/about/}{December 23, 2022} and \href{https://web.archive.org/web/20221230000607/https://georgeforny.com/about/}{December 30, 2022}. \cite{Frew2023Feb}}
    \label{fig:deletion-still}
\end{figure}

In our recent work \cite{frew-jcdl23}, Lesley Frew developed a prototype interface to allow users to search for deleted terms and phrases from a set of archived webpages, or \emph{mementos}. An example of this is shown in Figure \ref{fig:deletion-interface}, illustrating the results for searching for the deleted term ``jewish" from a set of mementos of politician George Santos' campaign website. The work also includes an interface that animates the deletion and addition of content between two mementos. Figure \ref{fig:deletion-still} shows a still image from the tool. The animated GIF and more examples are available in Lesley's blog post \cite{Frew2023Feb}.

\section{Validating Claims About Past Statements}
Web archives have also been used to hold public figures to account for what they have posted in the past. 
In 2018, Michael Nelson used multiple web archives, including the Library of Congress web archive,\footnote{\url{https://www.loc.gov/web-archives/}} to verify screenshots posted on Twitter of controversial content from MSNBC host Joy Ann Reid's former blog. Reid and her lawyers at one point claimed that the screenshots were fabricated or that the Internet Archive's versions of her blog had been hacked \cite{mediaite-joyreid}. Nelson \cite{Nelson-blog2018} described the process of using the Memento Time Travel service\footnote{\url{http://timetravel.mementoweb.org/}} and the Memgator tool,\footnote{\url{https://github.com/oduwsdl/MemGator}} developed at ODU, to track down mementos of Reid's blog in multiple web archives to corroborate the posted screenshots.

\begin{figure}
    \centering
    \includegraphics[width=0.45\textwidth, trim=65 110 70 0, clip]{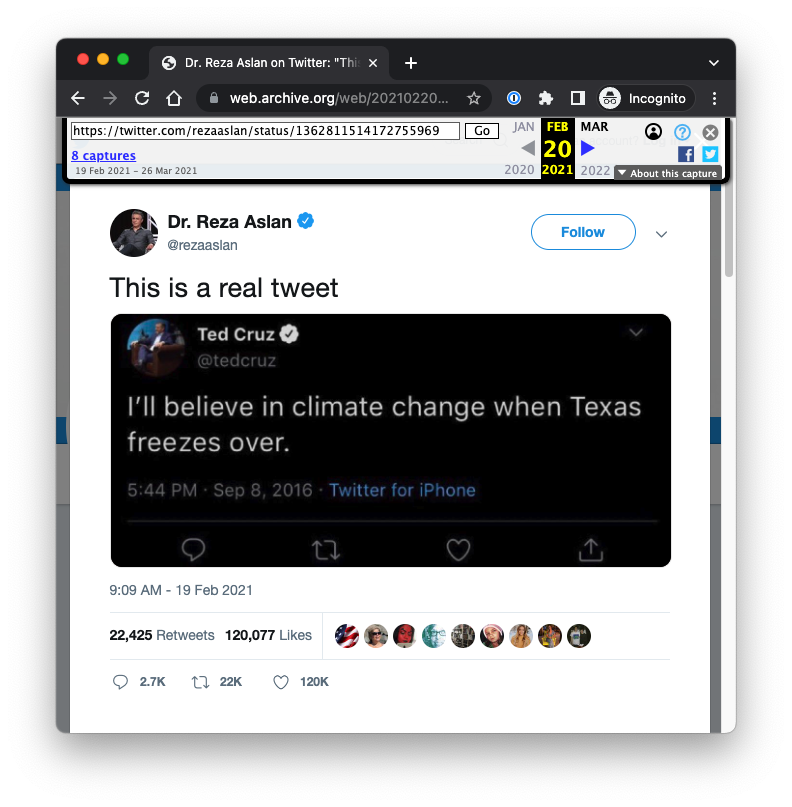}
    \caption{Archived version of a now-deleted tweet by Reza Aslan (@rezaaslan) sharing a fake tweet that was purportedly by US Senator Ted Cruz. The tweet sharing this fake was liked over 120,000 times and retweeted over 22,000 times. \cite{bradford-2022}}
    \label{fig:fake-tweet}
\end{figure}

This highlights one of the advantages of archiving controversial webpages rather than just taking screenshots. We all know that screenshots can be altered, but mementos stored in multiple independent web archives would be much harder to manipulate. Nelson's investigation spurred our current ``Did They Really Tweet That?'' project,\footnote{\url{https://twitter.com/phonedude\_mln/status/1485704384222830593}} funded by the US Department of Education, to use the live Web and web archives to help distinguish between fake attribution of social media screenshots and screenshots of real posts. Figure \ref{fig:fake-tweet} shows an example of a fake tweet \cite{cruz-fact-check} that was attributed to US Senator Ted Cruz.
As a part of this project, Tarannum Zaki has been building a dataset \cite{zaki-blog} of screenshots of tweets, both real and fake, that have been shared on social media. Her continuing work \cite{zaki-2023} is to build a system that could take a screenshot of a tweet and report the likelihood that it was an actual tweet. Initial work on this was performed as one of the projects in our 2022 NSF REU Site in Disinformation Detection and Analytics \cite{reu-blog}. In that project, Caleb Bradford built a tool that takes the text of a tweet and searches Politwoops, Reuters.com, and Snopes.com to determine if the tweet had been reported as fake \cite{bradford-2022}.

\section{Studying Archived Social Media}

\begin{figure}
    \centering
    \includegraphics[width=\textwidth]{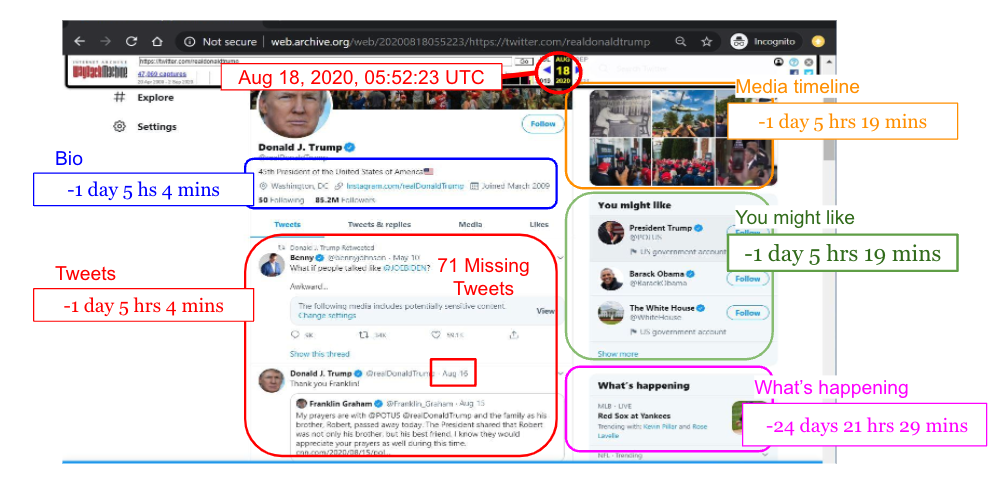}
    \caption{Annotated memento of @realDonaldTrump's account page from August 18, 2020, showing the temporal differences in the different sections of the page and highlighting that 71 tweets were missing from this memento. \cite{garg-jcdl21}}
    \label{fig:trump-temporal}
\end{figure}

With many politicians and other public figures using social media for communication, archived social media posts have become important objects of study. Because tweets from banned accounts are not available on the live Web, during former President Donald Trump's ban from Twitter between January 2021--November 2022, archives of his tweets became important public records \cite{Kriesberg2022Nov}.  This raises the issue of the quality of captures of social media in web archives. In our 
previous work \cite{garg-jcdl21}, Kritika Garg and Himarsha Jayanetti highlighted some of the issues related to replaying tweets from web archives. This included problems caused for crawlers by Twitter forcing the use of their new user interface (UI) in June 2020 \cite{garg-twitterblog-jul2020}, temporal violations upon replay with components of Twitter's new UI loading from different archived datetimes \cite{Jayanetti-twitterblog-nov2020}, and differences in how Twitter's fact check and violation labels replay in mementos between the old Twitter UI and the new Twitter UI \cite{garg-twitterblog-dec2020}.  Together these issues with archiving and replaying tweets from web archives could make it difficult to verify what might have appeared on live Twitter, especially for accounts that have been banned. For example, Figure \ref{fig:trump-temporal} shows how each section in an archived Twitter account page could be replayed from a different datetime. In particular, in this example, there are 71 missing tweets in the timeline because the tweet content comes from one day earlier than the datetime of the main memento.

\begin{figure}
    \centering
    \includegraphics[width=0.6\textwidth, trim=65 110 60 0, clip]{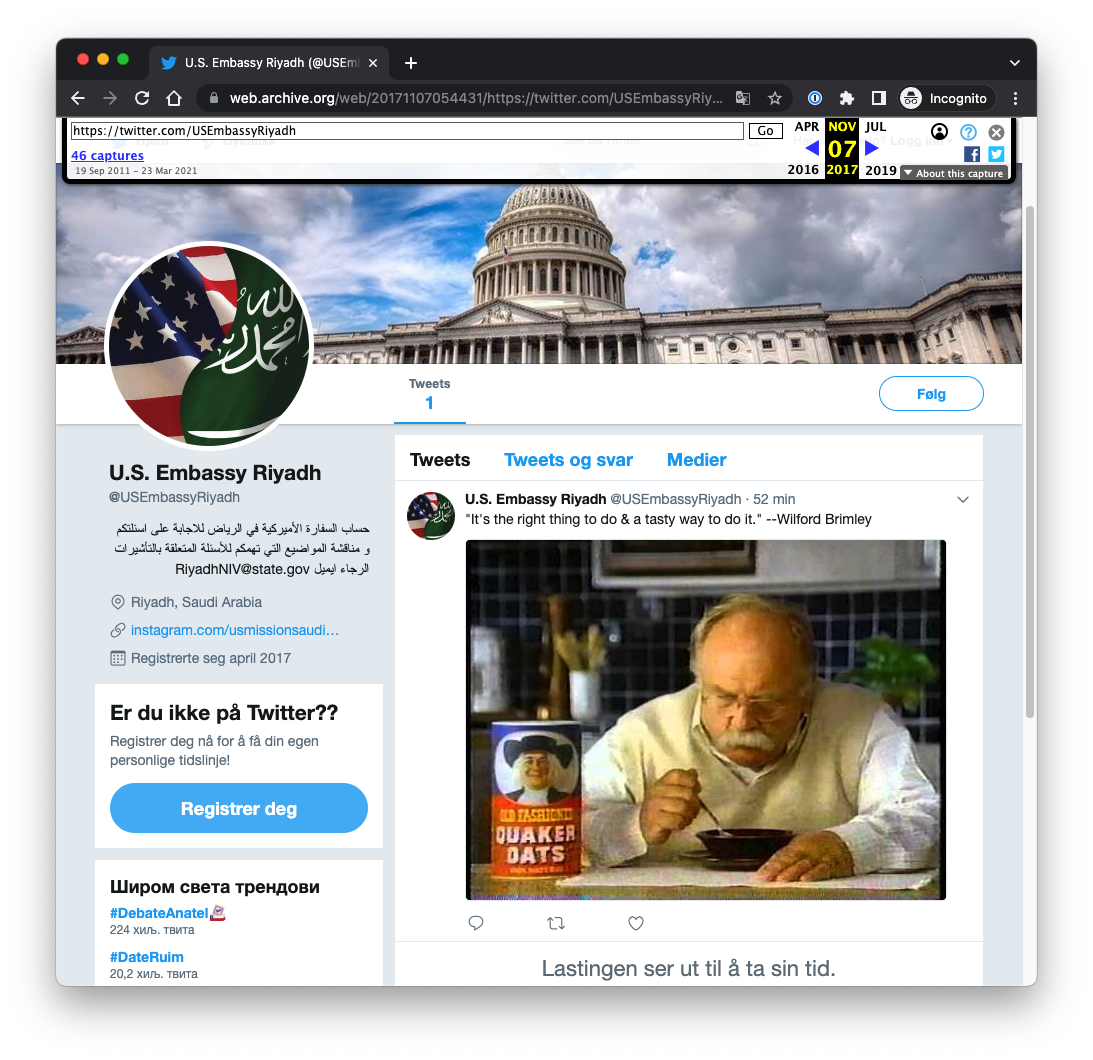}
    \caption{Archived tweet demonstrating the vulnerability of allowing dormant screen names of government agencies to be taken over by other users. \cite{littman-blog2017-2}}
    \label{fig:embassy-oatmeal}
\end{figure}

In 2017, while at George Washington University Libraries, Justin Littman noticed that several US government Twitter accounts that were listed in the US Digital Registry\footnote{\url{https://digital.gov/services/u-s-digital-registry/}} had been suspended by Twitter \cite{littman-blog2017-1}. He used the Wayback Machine to discover that after the accounts had been deleted, their screen names had apparently been taken over by other users \cite{littman-blog2017-2}.  Because the Wayback Machine indexes a webpage based on its URL and Twitter account URLs are based on screen names, mementos from both versions of the account would appear together in the Wayback Machine's list of mementos for the account.\footnote{\url{https://web.archive.org/web/20170601000000*/https://twitter.com/USEmbassyRiyadh}} Littman even demonstrated this by briefly taking over the @USEmbassyRiyadh account and pushing a tweet\footnote{\url{https://web.archive.org/web/20171107054431/https://twitter.com/USEmbassyRiyadh}} with a picture of actor Wilford Brimley into the Internet Archive (Figure \ref{fig:embassy-oatmeal}). 

\begin{figure}
    \centering
    \includegraphics[width=\textwidth]{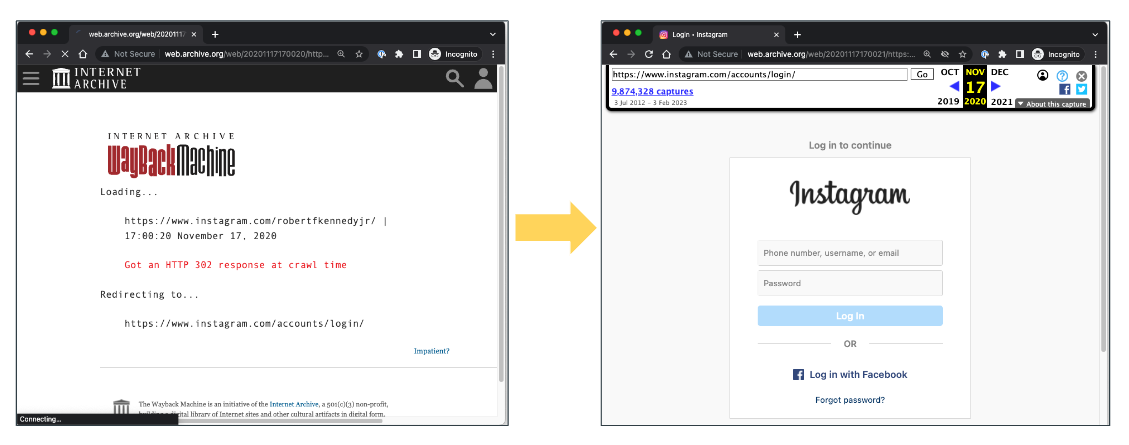}
    \caption{Many of the mementos of Instagram account pages redirect to Instagram's login page.}
    \label{fig:insta-redirect}
\end{figure}

\begin{figure}
    \centering
    \includegraphics[width=0.6\textwidth]{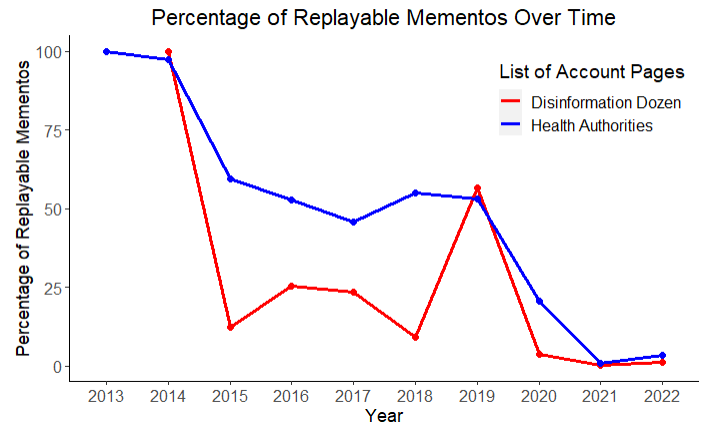}
    \caption{The percentage of replayable mementos of Instagram account pages has been decreasing over time, for both the Disinformation Dozen and for certain health authority accounts \cite{bragg-jcdl23}}
    \label{fig:replayable-over-time}
\end{figure}

Our research group has also explored how well (or not) Instagram has been archived,  using Katy Perry's popular account as a proxy for all of Instagram. Himarsha Jayanetti found that although Katy Perry is the 20th most popular person on Instagram, only about 1/3 of her posts are archived in public web archives \cite{jayanetti-instagram-blog}. Our interest in Instagram was piqued by the ``The Tactics \& Tropes of the Internet Research Agency'' report \cite{ira-report} that highlighted the IRA's heavy use of Instagram in their disinformation campaign. Instagram shut down their public API in 2018 \cite{techcrunch-instagramapi}, limiting the ability of researchers to study the platform, and thus the platform is much less studied than either Facebook or Twitter. In another one of the projects from our 2022 NSF REU Site in Disinformation Detection and Analytics, Haley Bragg analyzed \cite{bragg-reu} Instagram profile pages from the ``Disinformation Dozen'' \cite{disinfo-dozen} and found that 96\% of the mementos of those profile pages just redirected to the Instagram login page (Figure \ref{fig:insta-redirect}) \cite{bragg-jcdl23}, so the profile page content was not actually captured. 
She also found that the percentage of these mementos that are ``replayable'', meaning that they did not redirect to the Instagram login page, has decreased over time (Figure \ref{fig:replayable-over-time}). Many of these accounts have been banned from Instagram, so archived versions of their accounts are essential to study what they had been posting. This is a topic of ongoing research in our group. 

One of the issues with archiving social media is that traditional web archive crawlers, like the Internet Archive's Heritrix crawler, do not execute JavaScript while archiving a webpage, which may cause the archive to miss important resources loaded by JavaScript \cite{brunelle-ijdl16, weigle-jcdl23}. Browser-based crawlers, which typically use headless browsers, do execute JavaScript and often can create more complete mementos, but they run much more slowly \cite{brunelle-jcdl17}  and are not suitable for web archiving at scale. The Internet Archive has developed the Brozzler\footnote{\url{https://github.com/internetarchive/brozzler}} browser-based crawler and uses it to archive certain webpages, including those nominated using the Save Page Now\footnote{\url{https://web.archive.org/save/}} tool \cite{spn-brozzler}. The Webrecorder project\footnote{\url{https://webrecorder.net/}} has developed several tools\footnote{\url{https://webrecorder.net/tools}} leveraging browser-based crawlers, including ArchiveWeb.page\footnote{\url{https://archiveweb.page/guide}} and Browsertrix.\footnote{\url{https://github.com/webrecorder/browsertrix-crawler}} Other tools for archiving social media include Social Feed Manager\footnote{\url{https://gwu-libraries.github.io/sfm-ui/}} and twarc,\footnote{\url{https://github.com/DocNow/twarc}} which use APIs provided by the social media platforms to save social media posts and metadata. Unlike the crawlers mentioned above, they do not preserve the posts as they are displayed on the web.

The study of disinformation on social media often starts with a search for posts that contain certain keywords or hashtags. Kate Starbird's research group has used this technique to study tweets about alternative narratives regarding the 2010 Deepwater Horizon oil spill \cite{Starbird-medium16} (e.g., posts containing the hashtag ``\#oilspill''), mass shooting events \cite{starbird-icwsm17} (e.g., posts containing both terms about shootings and conspiracy-related terms such as ``false flag'', ``crisis actor'', ``staged'', ``hoax'') and the UN response in Syria \cite{Starbird-medium18} (e.g., posts containing the terms "white helmet" or "whitehelmet"). After gathering tweets with these terms that also contained URLs, they analyzed what domains were being linked to and often were able to identify groups of websites that formed an alternative media ecosystem \cite{starbird-icwsm18}.  In one study \cite{Starbird-medium16}, Starbird noted that they used the Wayback Machine to examine linked webpages that had changed or disappeared since the tweets were initially collected.

Often these initial investigations lead researchers to study individual accounts that are sharing disinformation, looking at who they are following, who is following them, whose posts they are sharing, who is sharing their posts. Amelia Acker \cite{acker-datacraft18} studied disinformation by examining the social media metadata (follower counts, etc.) of particular accounts that are known to share disinformation. Acker considers case studies including politicians on Instagram, US government Twitter accounts, and Facebook ads published by the Russian Internet Research Agency (IRA). The goal is to study the metadata of known manipulators to learn their techniques. Acker used web archives as a way of looking at social media in the past, including to examine the growth of an account's follower count. 

\begin{figure}[hbt!]
    \centering
    \includegraphics[width=0.75\textwidth]{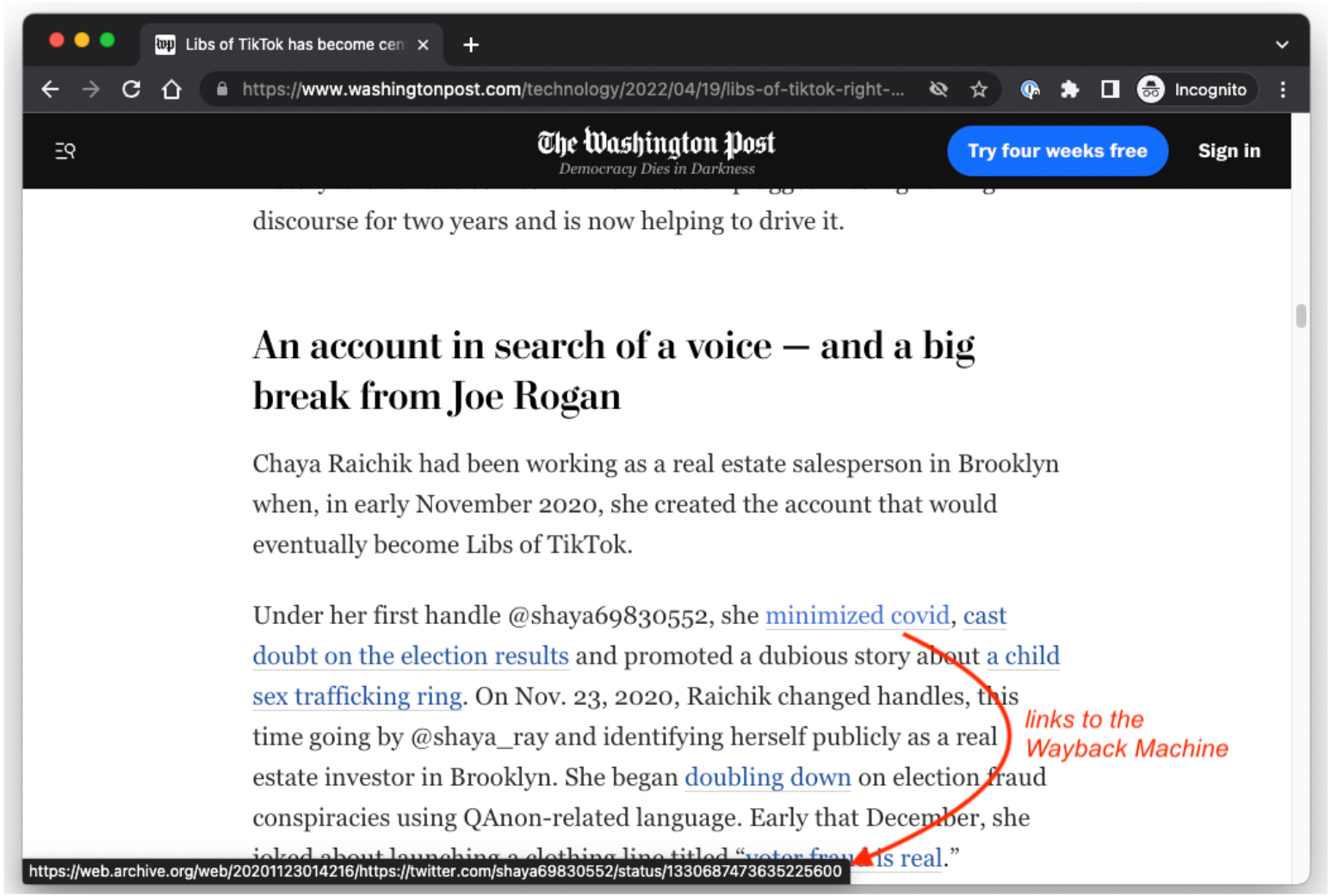}
    \caption{Annotated screenshot of Taylor Lorenz's article in the \emph{Washington Post} showing a link to a tweet archived at the Internet Archive.}
    \label{fig:lorenz-wapo}
\end{figure}

\begin{figure}
    \centering
    \includegraphics[width=0.75\textwidth]{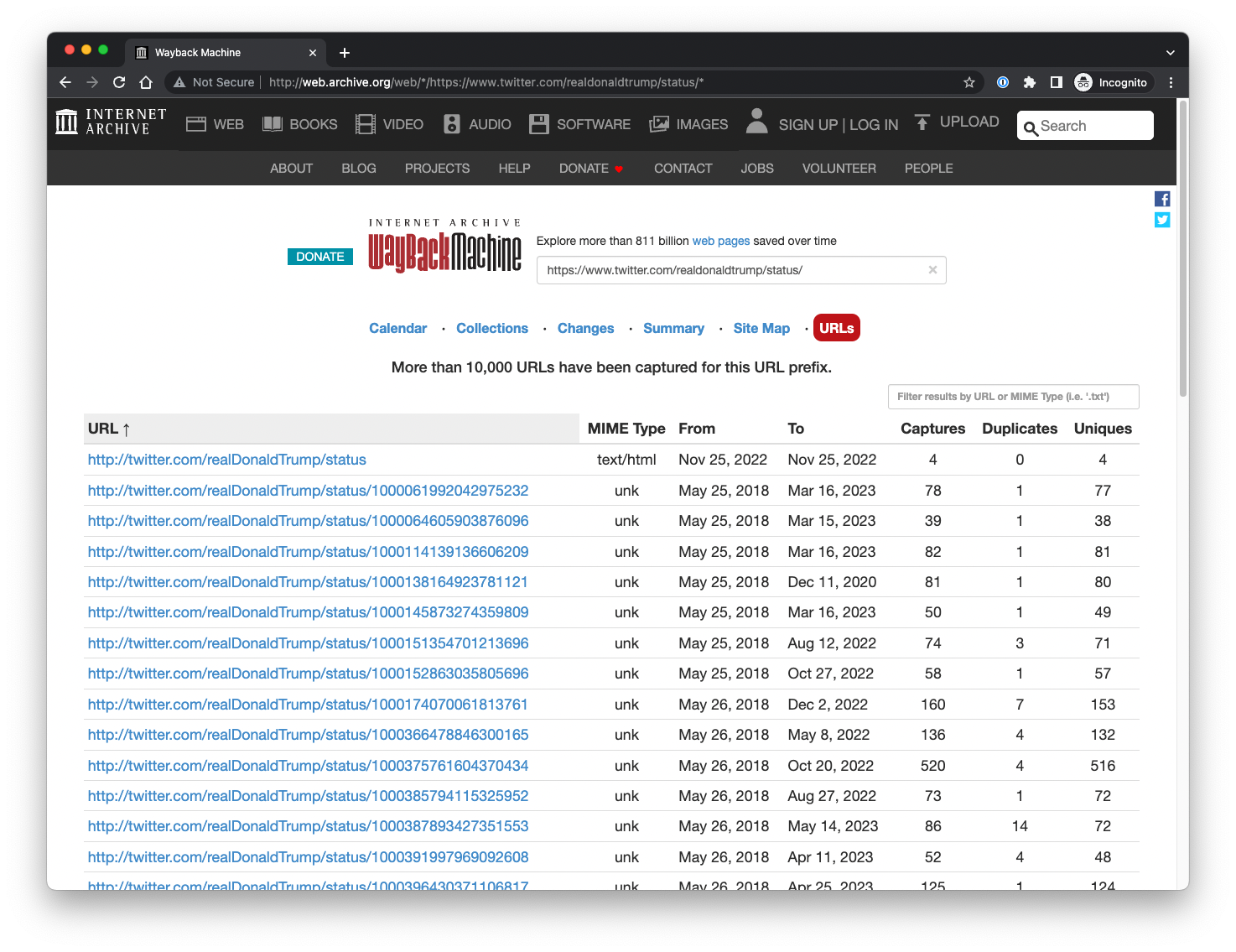}
    \caption{Example of using Wayback's CDX interface to find all archived tweets for a particular user (@realdonaldtrump), \url{http://web.archive.org/web/*/https://www.twitter.com/realdonaldtrump/status/* }}
    \label{fig:twitter-cdx}
\end{figure}

Ahmer Arif and others from Kate Starbird's research group used the Wayback Machine to analyze tweets and profile pages of accounts that had been suspended by Twitter \cite{arif-blm}. A similar technique was used by Taylor Lorenz in her reporting on tweets by the then-suspended ``Libs of TikTok'' account. Her article in the \emph{Washington Post} \cite{Lorenz-libsoftiktok}  is filled with links to tweets archived by the Wayback Machine (Figure \ref{fig:lorenz-wapo}) and is a great example of using web archives for evidence.
Related to these investigations, the Wayback Machine provides an interface that can list all mementos in the Internet Archive that start with a given prefix. Since the URLs of tweets have a standard format (https://www.twitter.com/\emph{username}/status/\emph{tweetid}), we can use this interface to find all of the tweets from a given username that are archived at the Internet Archive.  Figure \ref{fig:twitter-cdx} shows an example of this interface with a search for Donald Trump's (@realdonaldtrump) archived tweets.

\section{Studying Other Archived Content}

\begin{figure}
    \centering
    \includegraphics[width=0.75\textwidth, trim=0 250 0 0, clip]{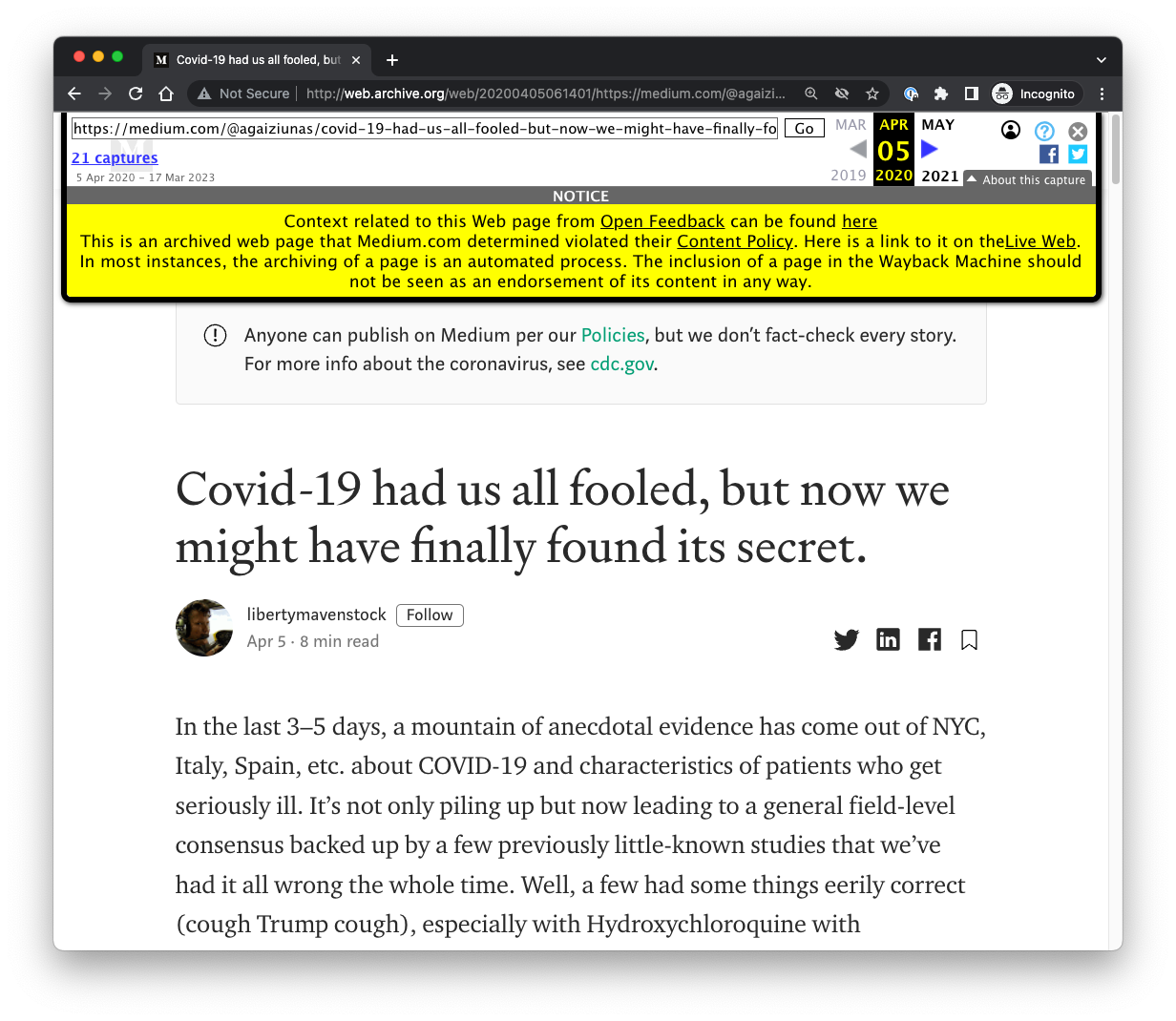}
    \caption{Example of webpage that had been removed from Medium for violating their content policies, but was being shared using its capture in the Wayback Machine. The Wayback Machine added a label to the replay of this known piece of disinformation.}
    \label{fig:labeled-memento}
\end{figure}

\paragraph{Known Disinformation.} 
Acker considers what web archives might hold that we have not yet discovered, ``Our web archives are now filled with examples of manipulation that were, at first, overlooked by platforms'' \cite{acker-datacraft18}. Unfortunately, purveyors of disinformation also know about web archives.  Some articles and posts that had been removed from their original platforms were still being shared via links to the Wayback Machine \cite{acker-misinfo20}.  However, this has prompted the Internet Archive to begin labeling some pieces of known disinformation \cite{wayback-factcheck}, as shown in Figure \ref{fig:labeled-memento}. 

\begin{figure}
    \centering
    \includegraphics[width=0.6\textwidth, trim=0 140 0 0, clip]{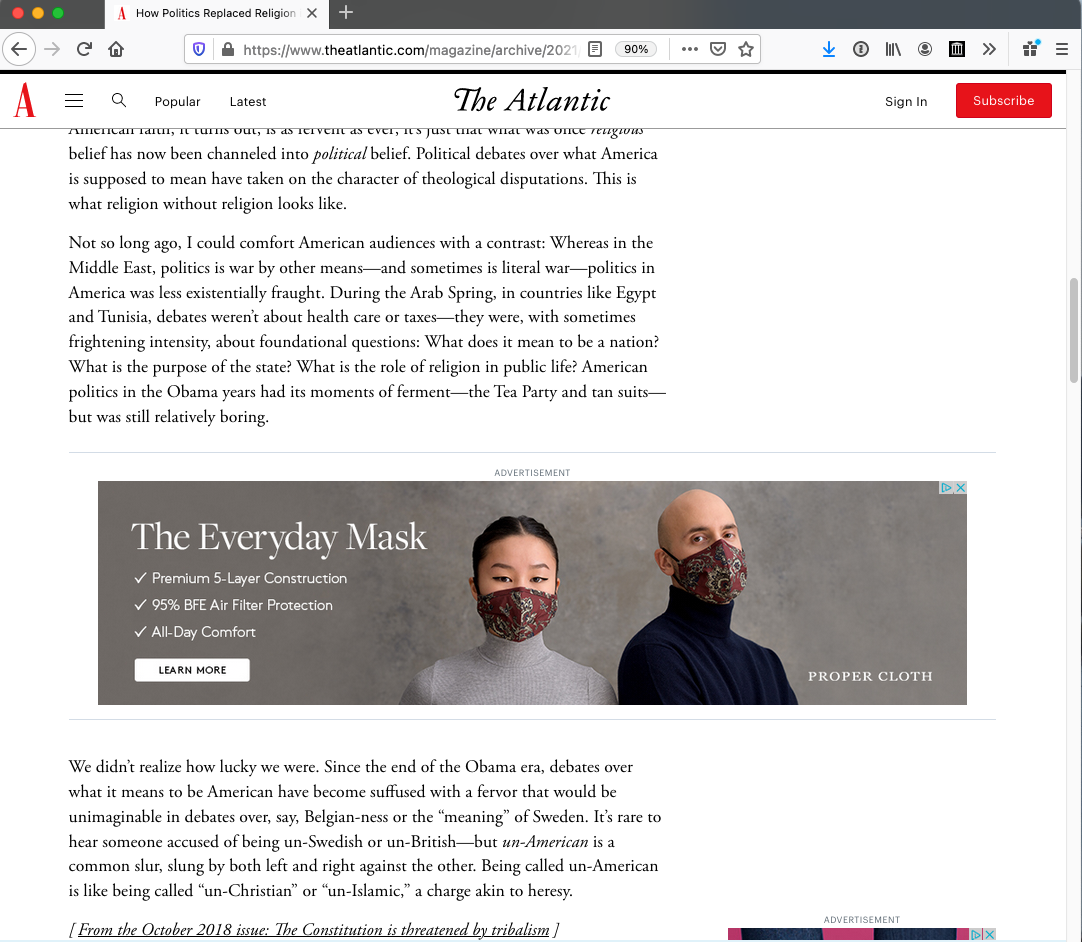}
    \caption{Example of an article containing an advertisement for COVID masks.}
    \label{fig:mask-ad}
\end{figure}

\paragraph{Advertisements.} 
Even pieces of the web that we might not consider important today, such as advertisements, may hold the key to understanding future disinformation campaigns, as suggested by Philip Howard, who wrote in a \emph{New York Times} editorial, ``Getting all ads, in real time, globally, into public archives is the next big step for strengthening democracy.'' \cite{howard-nytimes19} Our research group has long considered web advertisements to be important artifacts that should be archived, such as those that depicted the use of masks during the COVID-19 pandemic (Figure \ref{fig:mask-ad}).
Along with partners at Drexel University's College of Computing \& Informatics, we were recently awarded a grant from the IMLS for our project ``Saving Ads: Assessing and Improving Web Archives' Holdings of Online Advertisements'' \cite{imls-ads} to investigate this further.

\section{Conclusion}

This paper highlights just a few examples of how web archives have been used to investigate possible disinformation, by examining how webpages have been changed, corroborating or refuting posted screenshots, studying archived social media posts, and investigating metadata and past behavior of particular social media accounts. One of our goals in studying web archives and how people use them is to be able to provide analysis and develop tools and techniques that can help promote the use of web archives and provide more confidence in them.


\bibliographystyle{ACM-Reference-Format}
\bibliography{refs}

\end{document}